\documentclass{acm_proc_article-sp}
\usepackage{amsmath,graphicx}
\usepackage{subfigure, multicol}

\usepackage{amsmath,bm}  

\begin{document}

\title{Context-aware Ensemble of Multifaceted Factorization Models for Recommendation Prediction in Social Networks}


%
%
%
%
%

\numberofauthors{7} 
%

\author{
%
%
\alignauthor
Yunwen Chen\titlenote{Team leader and corresponding author.}\\
    \affaddr{Shanda Innovations}\\
    \email{kddchen@gmail.com}
\alignauthor
Zuotao Liu\\
\affaddr{Shanda Innovations}\\
\email{zuotao.liu@gmail.com}
\alignauthor
Daqi Ji\\
\affaddr{Shanda Innovations}\\
\email{duky2001@gmail.com}
\and
\alignauthor
Yingwei Xin\\
\affaddr{Shanda Innovations}\\
\email{xinyingwei@gmail.com}
\alignauthor
Wenguang Wang\\
\affaddr{Shanda Innovations}\\
\email{svd.wang@gmail.com}
\and
\alignauthor
Lu Yao\\
\affaddr{Shanda Innovations}\\
\email{luyao.2013@gmail.com}
\alignauthor
Yi Zou\\
\affaddr{Shanda Innovations}\\
\email{vodkajoey@gmail.com}
}

\maketitle
\begin{abstract}
This paper describes the solution of Shanda Innovations team to Task 1 of KDD-Cup 2012. A novel approach called Multifaceted Factorization Models is proposed to incorporate a great variety of features in social networks. Social relationships and actions between users are integrated as implicit feedbacks to improve the recommendation accuracy. Keywords, tags, profiles, time and some other features are also utilized for modeling user interests. In addition, user behaviors are modeled from the durations of recommendation records. A context-aware ensemble framework is then applied to combine multiple predictors and produce final recommendation results. The proposed approach obtained $0.43959$ (public score)/$0.41874$(private score) on the testing dataset, which achieved the 2nd place in the KDD-Cup competition. 
\end{abstract}

\category{H3.3}{Information Systems}{Information Search and  \\ Retrieval-information Filtering}



\section{Introduction}
Social Networking Services (SNS) have gain tremendous popularity in recent years, and voluminous information is generated from social networks every day. It is desirable to build an intelligent recommender system to identify what interests users efficiently\cite{DBLP:journals/survey}. The task\footnote{http://www.kddcup2012.org} of KDD-Cup 2012 Track 1 is to develop such a system which aims to capture users' interests,  find out the items that fit to users' taste and  most likely to be followed\cite{KDDCUP2012}. The datasets provided by Tencent Weibo, one of the largest social networking website in China is made up of 2,320,895 users, 6,095 items, 73,209,277 training records, and 34,910,937 testing records, relatively larger than other publicly released datasets. Besides, it provides richer information in multiple domains, such as user profiles, item categories, keywords, and social graph. Timestamps for recommendations are also given for performing session analysis. For each user in the testing dataset, an ordered list of the recommender results is demanded. Mean Average Precision (MAP)\cite{MuZhu04} is used to evaluate the results provided by 658 teams around the world.

Compared to traditional recommender problems, e.g.,the Netflix Prize\cite{DBLP:Netflix}, where the scores users rate movies are predicted, the settings of KDD-Cup 2012 appears more complex. Firstly, there are much richer features between users on social networking website. In the social graph, users can follow each other. Besides, three kind of actions, including ``comment'' (add comments to someone's tweet), ``retweet'' (repost a tweet and append some comments) and ``at'' (notify another user), can be taken between users. User profiles contain rich information, such as gender, age, category, keywords and tags. So models that are capable to integrate various features are required. Secondly, items that to be recommended are specific users, which can be a person, a group, or an organization. Compared to the items of traditional recommender systems, like books on Amazon or movies on Netflix, items on social network sites not only have profiles, but also have their behaviors and social relations. As a result, item modeling turns out more complicated. Thirdly, the training data in the social networks is quite noisy, and the cold-start problem also poses severe challenge due to the very limited information for a large number of users in testing dataset. It is demanding to have an effective preprocessing to cope with this challenge.

In this paper we present a novel approach called Context-aware Ensemble of Multifaceted Factorization Models. Various features are extracted from the training data and integrated in the proposed models. A two stage training framework and a context-aware ensemble method are introduced, which helped us to gain a higher accuracy. We also give a brief introduction to the session analysis method and the supplement strategy that we used in the competition to improve the quality of training data.

The rest of the paper is organized as follows. Section 2 will introduce preliminary of our methods. Section 3 presents the preprocess method we used. In Section 4, we will propose Multifaceted Factorization Models, which is adopted in the final solution. A context-aware ensemble and user behavior modeling methods are proposed in Section 5. Experimental results are given in Section 6 and conclusions and future works are given in Section 7.

\section{Preliminary}
In this section, we will first give a concise overview of the latent factor model, which forms the foundation of our solution. We then present some of its recent extensions specifically designed for collaborative filtering problems. Some preprocessing methods that we used are also introduced.

Before formally introducing the models, we first define our notational conventions. We denote $U$ as the user set and $I$ as the item set. Then we have the number of $|U|$ users and $|I|$ items, where function $|\cdot|$ indicates the quantity of members in a set. Special indexing letters are reserved for distinguishing users from items: for user $u$ and for item $i$. A binary rating $r_{ui}$ indicates the preference by user $u$ for item $i$, where value $r_{ui}=1$ means accepting the item and $0$ means rejecting. The ratings are arranged in a $|U|*|I|$ matrix $R=r_{ui}$. In the matrix, some cells have values from the training data, we use $\Omega$ to denote the set of $(u, i, t)$ records for which ratings are observed at time $t$. We denote $\Omega^{+}$ as all positive samples $\left(r_{ui} = 1\right)$ and $\Omega^{-}$ as negative sample set $\left(r_{ui} = 0\right)$.

\subsection{Latent Factor Model}
Recently, latent factor models comprise an alternative approach to collaborative filtering with the more holistic goal to uncover latent features that explain the relation between users and items\cite{NIPS07}. Because of its attractive performance, availability and scalability\cite{DBLP:journals/MF09}\cite{SVD07}, Singular Value Decomposition(SVD) is one of the most popular models. In this paper, we denote $\hat{r}_{ui}$ as the prediction of user $u$'s probability to follow item $i$. The baseline predictor of SVD with user bias and item bias is defined as: 
\begin{equation}
\hat{r}_{ui}=f\left(  b_{ui} + \bm{q}_{i}^{T}\bm{p}_{u} \right)
\label{E:basic-rui}
\end{equation}
where $f(\cdot)$ is a warping function used to map real values to certain ranges. $\bm{p}_{u}$ is the $d$-dimensional user feature vector and $\bm{p}_{u}$ is the item feature vector. $d$ is a parameter that set beforehand.  Bias factor  $b_{ui}$ is defined as:
\begin{equation}
b_{ui}=\mu + b_{u} + b_{i}
\label{E:basic-bui}
\end{equation}
In this equation, parameter $\mu$ denotes the overall average rating. $b_{u}$ and $b_{i}$ indicate the deviations of user $u$ and item $i$. These parameters can be learnt by solving the following regularized least squares problem:
\begin{multline}
\underset{\bm{p}_{*},\bm{q}_{*},b_{*}}{\operatorname{min}}
\sum_{(u,i)\in K}
\left({r}_{ui} - f\left( b_{ui} + \bm{q}_{i}^{T}\bm{p}_{u}\right) \right)^{2} \\
+ \lambda_{1}\left\|{\bm{p}_{u}}\right\|^{2}+\lambda_{2}\left\|{\bm{q}_{i}}\right\|^{2} +\lambda_{3} b^{2}_{u} + \lambda_{4} b^{2}_{i} 
\label{E:basic-loss-func}
\end{multline}

The first term in the above equation strives to find $b_{u}$, $b_{i}$, $\bm{p}_{u}$, and $\bm{q}_{i}$ that fit the given ratings. The others are regularization terms with parameters $\lambda_{1}$, $\lambda_{2}$, $\lambda_{3}$ and $\lambda_{4}$ to  avoid over-fitting.

\subsection{Model Learning Approach}
Stochastic Gradient Descent algorithm (SGD)\cite{DBLP:journals/MF09} is a very effective approach for solving the optimization problems(Eq.\ref{E:basic-loss-func}). It loops through known ratings randomly in set $\Omega$ and takes a small gradient descent step on the relevant parameters along the direction that minimizes the error on each rating. Let us denote the prediction error as $e_{ui}$. The SGD updates the parameters based on the following equations:
\begin{gather*}
b_{u} \leftarrow b_{u} + \eta \left(e_{ui} - \lambda_{1} b_{u} \right) \\
b_{i} \leftarrow b_{i} + \eta \left(e_{ui} - \lambda_{2} b_{i} \right) \\
\bm{q}_{i} \leftarrow \bm{q}_{i} + \eta \left(e_{ui}\bm{p}_{u}-\lambda_{3}\bm{q}_{i}\right) \\
\bm{p}_{u} \leftarrow \bm{p}_{u} + \eta \left(e_{ui}\bm{p}_{u} -\lambda_{4}\bm{p}_{u}\right) 
\label{E:sgd}
\end{gather*}
where parameter $\eta$ indicates the learning rate, $\lambda_{1}$, $\lambda_{2}$, $\lambda_{3}$, $\lambda_{4}$ are parameters that define the strength of regularization. The complexity of each iteration is linear in the number of ratings.

\subsection{Implicit Feedback}
Koren\cite{DBLP:conf/kdd/Koren08} indicated that the performance of recommendation systems can be improved significantly by accounting for implicit feedbacks. To solve the recommendation problem in social networks, high quality explicit feedbacks from users are usually limited, meanwhile, implicit feedbacks (e.g. the follow or be followed relations between users), which indirectly reflect opinion through observing user behavior, are more abundant and easier to obtain\cite{DBLP:conf/recsys/JamaliE10}. A conventional form of utilizing implicit information as user and item factor can be described in the following equation:
\begin{equation}
\hat{r}_{ui} = f\left( b_{ui} +\bm{q}_{i}^{T} ( \bm{p}_{u} + \sum_{k\in F(u)}\beta_{k}\bm{y}_{k} ) \right)
\label{E:implicit-feedback-rui}
\end{equation}
Here $F(u)$ indicates the set of user feedback records. $\bm{y}_{k}$ is a feature vector associated with the user feedback information.  $\beta_{k}$ is a parameter and usually set to $\left| F(u)\right|^{-1/2}$ in SVD++\cite{DBLP:conf/kdd/Koren08}.  In the competition, many kinds of features, such as: sns relations, user actions, keywords, tags, etc. were used in our solution as latent factors and experimental results proved their effectiveness.

\section{Preprocess}
This section introduces the procedure of data preprocessing and pairwise training in our solution

\subsection{Session Analysis for Data Filtering}
The training dataset consists of 73,209,277 binary ratings from 1,392,873 users on 4,710 items\cite{KDDCUP2012}. Among these ratings, there are 67,955,449 negative records and only 5,253,828 positive ones, which means  92.82\% of the training ratings are negative samples. After carefully analyzing the negative records, we found that not all the negative ratings imply that the users rejected to follow the recommended items. For example, users may be attracted by something interesting on the Weibo website, and during these sessions, the recommended items were omitted by users. So eliminating these "omitted" records from the training dataset $\Omega$ is a very important preprocessing. 

To remove those noise data, we adopt the following session analysis method. For each user $u$, we denote $t_{0}(u)$, $t_{1}(u)$, $t_{2}(u)$,\ldots, $t_{m}(u)$ as the timestamps of the recommended records in ascending order.  Then the time interval between two adjacent timestamp can be easily calculated:
\begin{equation}
 \Delta t_{s}(u)= t_{s+1}(u)- t_{s}(u)
\label{E:delta-t}
\end{equation}
 where $0<\Delta t_{s}(u)$, $0\le s<m$. When $\Delta t_{s}(u)<3600(s)$, we denote it as $\Delta \ddot{t}_{s}(u)$. The threshold for session slicing is then obtained: 
\begin{equation}
\tau\left(u\right)=\frac{1}{2}*\left(\tau_{0} + \frac{\sum_{s = 0}^{m} \Delta \ddot{t}_{s}(u)}{\left| \Delta \ddot{t}(u) \right|}\right)
\label{E:session1}
\end{equation}
 Parameter $\tau_{0}$ is set to 90 in our experiment. If $\Delta t_{s}(u) > \tau(u)$, the training records on time $t_{s+1}(u)$ and $t_{s}(u)$ will be seperated to different subsets (sessions) $\psi_{k}(u)$ and $\psi_{k+1}(u)$. We denote $\psi_{k}^{+}(u)$ as all positive samples in $\psi_{k}(u)$.

For each session $\psi_{k}(u)$ and records $r_{ui}\in\psi_{k}(u)$, their validity are carefully evaluated. Only those samples which can simultaneously satify the following three formulas are reserved as valid training data.
\begin{gather}
\sigma_{s}-\sigma_{-}\le \pi_{-} \\
\sigma_{+}-\sigma_{s} \le \pi_{+} \\
0 < \frac{\left|\psi^{+}_{k}(u)\right|}{\left|\psi_{k}(u)\right|} \le \epsilon
\label{E:session2}
\end{gather}
$\sigma_{s}$ denotes the timestamp index of a record within the session. $\sigma_{-}$ is the smallest index value of all the records $\sigma_{k}\in \psi_{k}^{+}(u)$, and $\sigma_{+}$ is the largest one as well.  $\pi_{-}$ , $\pi_{+}$ and $\epsilon$  are parameters on the effect of filtering. In our approach they are set to be 0, 3, 0.86, respectively. We picked out 7,594,443 negative records and 4,999,118 positive ones from the training dataset, and experimental results show that the proposed preprocessing effectively improves the prediction accuracy.

\subsection{Pairwise Training}
\begin{figure}
  \includegraphics[width=8.83cm]{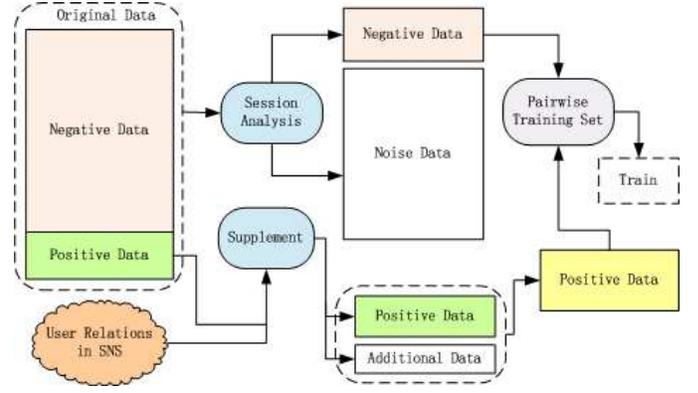}\\
  \caption{The procedure of data preprocessing and pairwise training}
  \label{fig-preprocess}
\end{figure}
Different from conventional recommender applications, where Mean Square Errors (MSE) is often used to evaluate the predicted results, Mean Average Precisions (MAP) is employed in the competition as the evaluation metric\cite{MuZhu04}.  For each candidate user in the testing dataset, the top-N items (N=3 in the competition) this user are most likely to follow are required, and average precisions of the top-N results are then calculated. MAP is the mean value of all these average precisions.  Since MAP is commonly used to judge the performance of different ranking algorithms, the KDD-Cup Track 1's task is more like a learning-to-rank problem\cite{TieyanLiu10}. In this case, we reformulate the task as a pairwise ranking job. For a given negative training sample $(u,i)$ and a positive one $(u,j)$, the ranking objective is to give higher prediction value $\hat{r}_{uj}$ over  $\hat{r}_{ui}$. The objective function for this pairwise training is:
\begin{equation}
g\left(\hat{r}_{ui}, \hat{r}_{uj}\right) =   \left(1 + exp\left(-\left( b_{j} - b_{i} + \left(\bm{q}_{j} - \bm{q}_{i}\right)^{T}\bm{p}_{u}\right) \right) \right)^{-1}
\label{E:basic-pairwise}
\end{equation}
Here we use sigmoid function  for value wrapping. For each negative sample $r_{ui}\in \Omega^{-}$, we randomly select a positive sample $r_{uj}$ in $\Omega^{+}$ to build a training pair ($r_{ui}$, $r_{uj}$). Similarly to Eq.(\ref{E:basic-rui}), SGD can be used here for  optimization. Note that $b_{u}$ is omitted because  user biases to item $i$ and $j$ are always the same.

\subsection{Supplement Training Data}
An ideal pairwise training requires a good balance between the num of negative and positive samples. However, in training dataset 94.9\% of the users have less positive samples than negative ones. To deal with this problem, a preprocessing of sample supplement is added in our approach. Firstly, the users who have a far smaller number of positive samples than the negative samples are chosen. Then for each user $u$, we denote $S(u)$ as the items that the user followed and $A(u)$ as those user have interactions ('at', 'retweet' or 'comment') with. The best item was selected by the following function, and added to positive training set $\Omega^{+}$.
\begin{displaymath}
\underset{i \in S(u) \cap  A(u) }{\operatorname{arg max}} \left( \xi_{0} N_{at}(u,i) + \xi_{1} N_{retweet}(u,i)  + \xi_{2} N_{comment}(u,i)  \right)
\end{displaymath}
here $N_{at}(u,i)$,$N_{retweet}(u,i)$,$N_{comment}(u,i)$ indicate the numbers of each kind of actions user $u$ gave to item $i$. Parameters $\xi_{0}$,$\xi_{1}$,$\xi_{2}$  are set to 2,0.2,1 respectively. 

Figure \ref{fig-preprocess} shows the procedure of data processing mentioned above. We filtered noise data through session analysis, and added  positive samples from social graph. Then built set $(\hat{r}_{ui}, \hat{r}_{uj})$ for pairwise training.

\section{Multifaceted Factorization Models}
This section describes our approach named Multifaceted Factorization Models (MFM) used for user interest prediction. The motivation of this approach is to integrate all kinds of valuable features in social networks to improve the recommendation effects. MFM extend the SVD function to:  
\begin{equation}
\hat{r}_{ui} = \ddot{b}_{ui} +  d_{ui} + \ddot{\bm{q}}_{i}^{T}\ddot{\bm{p}}_{u}
\label{E:MFM}
\end{equation}
where $\ddot{b}_{ui}$,$d_{ui}$, $\ddot{\bm{q}}_{i}$, $\ddot{\bm{p}}$ are defined in the following equations:
\begin{multline}
\ddot{b}_{ui} = \mu + b_{u} + b_{i} + b_{hour} + b_{day(i)} + b_{sec(i)} \\
+ b_{u,gender(i)} + b_{u,age(i)} + b_{gender(u),i} + b_{age(u),i}\\
+ b_{kw(u,i)} + b_{tag(u,i)} + b_{tweetnum(u)}
\label{E:MFM-bui}
\end{multline}
\begin{multline}
d_{ui} = \left|R^{k}(u,i)\right|^{-\frac{1}{2}}\sum_{j\in R^{k}(u,i)}\left(r_{ui} - 
\bar{r}_{uj}\right)w_{ij} \\
+ \left|N^{k}(u,i)\right|^{-\frac{1}{2}}\sum_{j\in N^{k}(u,i)}c_{ij}
\label{E:MFM-dui}
\end{multline}
\begin{gather}
\ddot{\bm{q}}_{i} = \bm{q}_{i} + \bm{z}_{day(i)} + \bm{z}_{sec(i)}
\label{E:MFM-q}
\end{gather}
\begin{multline}
\ddot{\bm{p}}_{u} = \bm{p}_{u} + \bm{y}_{age(u)}+\bm{y}_{age(u),gender(u)} + \bm{y}_{tweetnum(u)}\\
+ \left|S(u)\right|^{\alpha_{1}}\sum_{k\in S(u)} \bm{y}_k + \left|A(u)\right|^{\alpha_{2}}\sum_{l\in A(u)} \bm{y}_{l} \\
 + W(u,m)\sum_{m\in K(u)} \bm{y}_{kw(m)} + \left|T(u)\right|^{-\frac{1}{2}}\sum_{n\in T(u)} \bm{y}_{tag(n)}
\label{E:MFM-p}
\end{multline}
Detailed descriptions of these functions are given below.

\subsection{Date-Time Dependent Biases}
Temporal information is proved to be very effective for recommendation\cite{DBLP:CF/koren}\cite{KDDCUP2011:TianqiChen}. Firstly, users' action differs when time changes. For instance, in the day time a user who is busy with work may be less willing to follow and tweet others than in the evening, or a user's behavior may be vary  between weekdays and weekends. Secondly, items' popularities are changing. As a result, we incorporate temporal information to  $\ddot{b}_{ui}$ and  $\ddot{\bm{q}}_{i}$ in our solution.

Firstly, We designated three time factor as global biases and add them to $b_{ui}$ in Eq.(\ref{E:MFM-bui}).
\begin{equation}
b_{hour} = b_{Bin(t)} 
\end{equation}
\begin{equation}
b_{day(i)} = \frac{t-d^{-}}{d^{+}-d^{-}}b_{day(i)}^{-}+\frac{d^{+}-t}{d^{+}-d^{-}}b_{day(i)}^{+}
\label{E:DAY}
\end{equation}
\begin{equation}
b_{sec(i)} =  \frac{t-s^{-}}{s^{+}-s^{-}}b_{sec(i)}^{-}+\frac{s^{+}-t}{s^{+}-s^{-}}b_{sec(i)}^{+}
\label{E:SEC}
\end{equation}
where $t$ is the timestamp of a recommendation record. $Bin(t)$ denote the bin index associated with the hour (a number between 1 and 24). $d^{-}$ and $d^{+}$ indicate the begin time and end time of the entire dataset. $s^{-}$ and $s^{+}$ indicate the begin time and end time of a day. In our experiment they are set to 1318348785, 1322668798, 0, and 86400 respectively.  

Apart from the global biased, date and time factors can also be used as latent factors of items (shown in Eq.\ref{E:MFM-q}). Regardless of whether the items were followed or not,  the date and time factors of the recommender records can tell us about item's preferences. So we extended the item-factors vector $\bm{q}_{i}$ by adding $\bm{z}_{day(i)}$ and $\bm{z}_{sec(i)}$, which are defined by functions similar to linear Eq.(\ref{E:DAY}) and (\ref{E:SEC}). Note that we did not add time factors to $p_{u}$, because we found that temporal information for users is very sparse, and has little positive effect to the performance.
\subsection{Integrate Social Relationships}
A phenomenon observed from testing data is that 77.1\% users do not have any rating records in the training set. Therefore it is very difficult to predict these users' interests merely through explicit feedbacks.  Fortunately, due to social relationships, people are often influence in attitude, taste or actions by their peer groups. M. Jamali\cite{DBLP:conf/recsys/JamaliE10} and Ma\cite{Ma:2011:RSS:1935826.1935877} proved that  exploiting social network could alleviate the data sparse problem and improve the recommendation accuracy, especially for those cold start users.

In this section, two kinds of social relationships are introduced and used as the implicit feedbacks. The first and foremost implicit feedback is the user's SNS followships. Given a user $u$, we denote $S(u)$ as the set of users followed by $u$.  Similar to SVD++ approach \cite{DBLP:journals/sigkdd/BellK07}, we incorporated $S{u}$ factors to $\bm{p}_{u}$, which means that one user's interests can be influenced by the users he followed.The user latent vector $\bm{p}_{u}$ is extended to:
\begin{equation}
\bm{p}_{u}+\left|S(u)\right|^{\alpha_{1}}\sum_{k\in S(u)} \bm{y}_k 
\end{equation}
Besides, we also explored the user actions, including ``comment'', ``retweet'', and ``at''. If  user $u$ has an action to user $v$, there may exist some latent common interests between them. The factorization model is then updated to:
\begin{equation}
\bm{p}_{u}+\left|S(u)\right|^{\alpha_{1}}\sum_{k\in S(u)} \bm{y}_k +\left|A(u)\right|^{\alpha_{2}}\sum_{l\in A(u)} \bm{y}_{l} 
\end{equation}
$A(u)$ is the set of users $u$ had given actions to. $\alpha_{1}$  and $\alpha_{2}$ are parameters for  normalization. We set them to -0.4 and -0.5 in the experiments. 

It should be mentioned that a effective preprocess we did is to add each $u$ to $S(u)$, which means that user $u$ is a default follower of himself. Experimental result shows that it can improve the accuracy of recommendation.   
\begin{figure*}
  \centering
  \includegraphics[width=14.82cm]{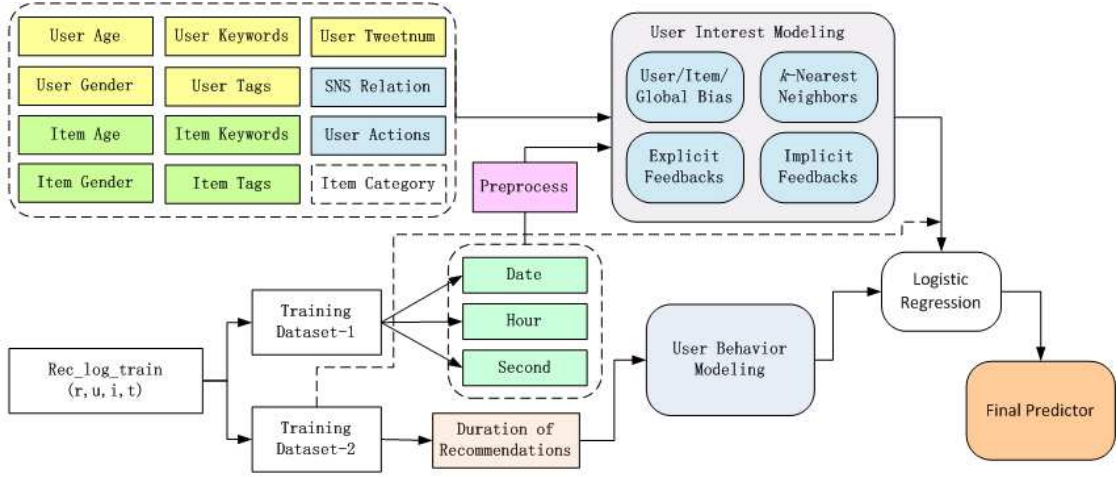}\\
  \caption{Context-aware ensemble of multifaceted factorization models}
    \label{Ensemble}
\end{figure*}

\subsection{Utilizing Keywords and Tags}
Keywords are extracted from the tweet/retweet/comment of users, or from their self-introductions. Each keyword $m$ has a weight $W(u,m)$ represents the relevance to user $u$. These weighs are pre-calculated and provided by competition orgnizers. While tags are selected by users to give a summary of the topics they are interested in. Keywords and tags use different vocabularies. Both  of them were used to explore users' latent interests. 

Firstly, if user $u$ share the same keyword or tag with item $i$, this keyword or tag can be seen as a feature of the common interests between $u$ and $i$. The deviation of this keyword can be trained through the explicit feedbacks. So we added global bias $b_{kw(u,i)}$ and $b_{tag(u,i)}$ to $\ddot{b}_{ui}$ in Eq.(\ref{E:MFM-bui}), where:
\begin{gather}
b_{kw(u,i)} = \sum_{m \in K(u)\cap K(i)}b_{m} \\
b_{tag(u,i)} = \sum_{n \in T(u)\cap T(i)}b_{n}
\end{gather}
Here $K(\cdot)$ denotes the keyword set of a given user and $T(\cdot)$ denotes the corresponding tag set.  

The second method to make use of keywords and tags is to consider them as latent factors. We believe that keywords and tags can be presented as parts of user-factors vector, so the following two terms were added:
\begin{equation}
W(u,m)\sum_{m\in K(u)} \bm{y}_{kw(m)} + \left|T(u)\right|^{-\frac{1}{2}}\sum_{n\in T(u)} \bm{y}_{tag(n)}
\end{equation}
We found that weights $W(u,m)$ of the keywords from tweets have already been well normalized. As for every user, $W(u,m)$ satisfy
\begin{equation}
\sum_{m\in K(u)}W(u,m)^2  \approx 1
\end{equation}
On the other hand, keywords from user descriptions (i.e. $weight=2.0$) have a lower quality, since they are extracted directly from the descriptions and without any postprocess.  So in our models we only utilized the former.

\subsection{$k$-Nearest Neighbors}
The $k$-Nearest Neighbor approach is also used to improve the results. As shown in Eq.(\ref{E:MFM-dui}), $R^{k}(u,i)$ is the intersection between user $u$'s explicit feedbacks and its $k$-nearest neighbors $D^{k}(u)$, and $N^{k}(u,i)$ is the intersection between implicit feedbacks and $D^{k}(u)$ . The implicit feedback set used here is sns relationship $S(u)$. $\bar{r}_{uj}$ is an average rating of $u$ pre-calculated. We used it as the approximation of ${r}_{uj}$. The item-item weights $w_{ij}$ and $c_{ij}$ represent the  adjustments that need to make to the predicted rating of item $i$, given a known rating $\bar{r}_{uj}$ of neighbor $j$. 

Tags and keywords were used to calculate the distance between items and find out the neighbors. 
We denote $dis(i,j)$ as the distance between items $i$, $j$, and calculate it through the following linear function:
\begin{equation}
dis(i,j) = \rho* dis_{kw}(i,j) + \left(1-\rho\right)*dis_{tag}(i,j)
\label{E:item-sim}
\end{equation}
where $dis_{tag}(i,j)$ and $dis_{kw}(i,j)$ are distances between two items in  vector space of tags and keywords respectively. $\rho$ is a paramter set to 0.6.  We mapped item $i$  to a vector $kw(i)$ with the keywords $m\in K(i)$ and weight $W(i,m)$. Cosine distance was then used to calculate the distance $dis_{kw}(i,j)$ betweent  two vectors $kw(i)$ and $kw(j)$. $dis_{tag}(i,j)$ was obtained in the same way.

\subsection{User and Item Profiles}
The profiles of users and items, such as gender and age, were proved to be very effective in modeling user interests. Users of comparable ages may share some common tastes. For example,  users born in 1960s may be more familiar with the acters populared a few decades ago. Gender is  another valuable feature for user interest modeling. An example is that  males generally are more interested in military topics than females. So we added four bias term $b_{u,gender(i)}$ and $b_{u,age(i)}$  to the prediction formula. They represent the users' preferences to particular groups of items, who have the same gender or age. While $b_{gender(u),i}$ and $b_{age(u),i}$ represent the derivation of how attractive the item to particular type of users. We denote user $u$'s age as $x$, and
\begin{gather}
age(u) = 
\begin{cases}
0, &\text{if $x < 1950$} \\
ceil((x - 1950)/2)+1, &\text{if $1950 \le x < 2004$} \\
28,  &\text{if $x \ge 2004$} \\
29, &\text{if $x$ is illegal.} 
\end{cases}
\end{gather}
We also integrate $y_{age(u)}$ and $y_{age(u),gender(u)}$ in Eq.(\ref{E:MFM-p}) to influnce user vector $p_{u}$, because age and gender can also be handled as implicit feedbacks.

Tweet number of users is a very minor profile. However, we still use it through add $b_{tweetnum(u)}$ in Eq.(\ref{E:MFM-bui}) and $y_{tweetnum(u)}$ in Eq.(\ref{E:MFM-p}). We wrap it to the range of [0, 15] by a log function beforehand. 

Some other features, including categories of items, are abandoned, because they were considered to be low quality and did not bring positive effect in our models.
\section{Context-aware Ensemble}
In this section, we first introduce the framework of our $2$-level ensemble method. Then detailed introductions to user behavior modeling and training process are given. Finally, a supervised regression used to combine different predictor results is also introduced. The overall view of the workflow is presented in Figure \ref{Ensemble}.

\subsection{Framework}
Shown in Figure \ref{Ensemble}, our solution can be divided into three core stages. The first stage is the user interest modeling. We extend the SVD model to the proposed multifaceted factorization models in order to incorporate various features in social networks. The second stage is user behavior modeling using durations of recommendations.  The user interest model and user behavior model is ensemble in the final stage. Logistic regression is used for a linear combination. During the ensemble, not only the predictor results of current timestamp is considered, but also the user's behaviors before and after are involved. In other words, the ensemble incorporates the ``context'' of recommendations. 

The data used in the framework is set as follows. We split the original training set into two parts: $\Phi_{1}$ is the records from first 23 days, and $\Phi_{2}$ is those from the next 7 days.  We firstly trained  the multifaceted factorization models on  $\Phi_{1}$, and made predictions on $\Phi_{2}$. Then we modeled user behaviors and trained our context-aware ensemble based on data $\Phi_{2}$. Lastly the whole training set was used to train the final predictor. After that, the final predictions were obtained on testing data.

\subsection{User Behavior Modeling}
The goal of the competition is to recommend up to three items that the user was most likely to follow. Yet there are two necessary conditions for the user to follow the item: 1) the user noticed the items recommended to him, 2) the item aroused his interests.   

As stated earlier, the factorization models were used to locate users' interests. So in this section we introduce the solution about predict whether the user noticed the recommendation area. We found that the durations of users on each recommendation are very valuable clues to model users' behaviors on social network website. For example, a user who was tweeting with friends and refreshing the page may have a shorter duration on the website than those who were reading the webpage carefully.

In Eq.(\ref{E:delta-t}) we already defined the time interval $\Delta t_{s}(u)$. Before using temporal information, we first spread $\Delta t_{s}(u)$ into discrete bins $\delta_{s}(u)$ to alleviate  data sparsity.
\begin{displaymath}
\delta_{s}(u) =
\begin{cases}
-1, &\text{if $ s < 0$}\\
k, &\text{if $0\le s < m$, and $\theta_{k}\le\Delta t_{s}(u)< \theta_{k+1}$ }\\
-1, &\text{if  $s \ge m$}
\end{cases}
\end{displaymath}
We set bin number to 16 in our solution. Parameters $\theta_{k}$ were assigned to have the bin well-distributed. $\delta_{s}(u)$ was used for modeling users' browsing habit in social networks. Let us denote $P(\delta_{s})$ as the probability that a user noticed the recommendation area when the duration was $\delta_{s}$. Then the training dataset $\Omega^{-}$ and $\Omega^{+}$ were used to calculate the probabilities.

In addition, users' behavior not only can be predicted through the duration of current webpage, but also can be analyzed according to the durations of the webpages before and after. For instance a user who hold a very short duration before and after, and took a long time on the current page, may have a higher probability to pay attention to the recommended items. Thus, we have the duration factor $\Gamma(r)$, where $r$ denotes the range of the context considered. The formulas of $\Gamma(r)$ used in our solution is listed below.
\begin{align}
&\Gamma(1) =  P(\delta_{s}) \\
&\Gamma(2) =  P(\delta_{s-1}, \delta_{s}) \\
&\Gamma(3) =  P(\delta_{s-1}, \delta_{s}, \delta_{s+1}) \\
&\Gamma(4) =  P(\delta_{s-2}, \delta_{s-1}, \delta_{s}) +P(\delta_{s-1}, \delta_{s}, \delta_{s+1}) \\
&\Gamma(5) =  P(\delta_{s-2}, \delta_{s-1}, \delta_{s}) +P(\delta_{s-1}, \delta_{s}, \delta_{s+1}) +P(\delta_{s}, \delta_{s+1}, \delta_{s+2}) 
\end{align}
$\Gamma(r)$ will be blended  with  multifaceted factorization models through supervised regression method.

\begin{table}
\centering
\caption{Performances of Different Multifaceted Factorization Models on level 1}
\begin{tabular}{l|p{0.365\textwidth}|p{0.09\textwidth}} \hline
ID	& Description & MAP \\ \hline
1	& basic matrix factorization($d$=40,$\eta$ =0.001) & 0.3442 \\ \hline
2	& 1+pairwise training & 0.3458 \\ \hline
3	& 2+preprocess of the training data & 0.3495 \\ \hline 
4	& 3+sns as implicit feedback & 0.3688 \\ \hline 
5	& 4+action as implicit feedback & 0.3701 \\ \hline
6	& 5+day as bias and item factor & 0.3721 \\ \hline
7	& 6+second as bias and item factor & 0.3728 \\ \hline
8	& 7+age\&gender as bias and user factor & 0.3854 \\ \hline
9	& 8+user tags as bias and user factor & 0.3863 \\ \hline
10	& 9+user keywords as bias and user factor & 0.3887 \\ \hline
11	& 10+tweetnum as bias and user factor & 0.3892 \\ \hline
12	& optimize parameters in 11($d$=100,$\eta$=0.00025) & 0.3901 \\ \hline
13	& 12+nearest neighbors & 0.3907 \\ \hline
14	& 13+supplymental training data & 0.3911 \\ \hline
\end{tabular}
\label{tab1}
\end{table}

\subsection{Supervised Regression}
It has been proved that combination of different predictors can lead to significant performance improvement over individual algorithms\cite{DBLP:ENSEMBLE}. Ensemble a set of predictors' results is a standard regression problem. In our solution, we choose logistic regression as the tool to deal with the problem. The big advantage of logistic regression is clearly the speed. We can  generate our final predictor efficiently. 

Firstly, we trained the multifaceted factorization models independently based on training dataset  $\Phi_{1}$, and used it to predict the ratings in set $\Phi_{2}$. The predictions from user behavior models were also obtained. Then we got a total of $M$ result vector $F_i$, where $M$ is the number of individual predictors, and  $F_i$ is a $N$-dimensional vector.  $N$ equals the  size of validation set $\left|\Phi_{2}\right|$. 
\begin{equation}
Z=W_{0} + \sum_{i=1}^{m}W_{i}*{F_i}
\end{equation}
where $W_i$ is the coefficient for different individual models. $Z$ is the prediction vector $Z = \left(z_0, z_1, ....,z_{N-1}\right)$. Then prediction value of the $i$-th record is 
\begin{equation}
\hat{r_i}=\frac{1}{1 + exp{(-1*z_i)}} 
\end{equation}
Parameters $W_i$ can be easily obtained through the logistic regression approach. At last the proposed multifaceted factorization models were trained again using the same initialization parameters on the entire training set, and  then the final prediction was obtained on the testing dataset.

\section{Experiments}
This section presents our experimental results. Detail information of the multifaceted factorization models is given in Table \ref{tab1}. We can conclude that user sns relations bring significant benefits to the model. User profiles, such as age and gender, are also very helpful. Since the cold start challenge is grave in the task, these features are very promising supplement to locate interests of  inactive users.

Table \ref{tab2} shows the  performance of ensembles, where level-1 models with \( \Gamma(5) \) ensemble gives the best performance. Durations of the recommendations show great effectiveness to predict users' response. The utilization of recommendation context also gives improvement. Although more context information brings better performance, the risk of overfitting also grows, so we limit the range of context to 5.

\begin{table}
\centering
\caption{Performance of Context-aware Ensemble on level 2}
\begin{tabular}{c|p{0.32\textwidth}|p{0.1\textwidth}} \hline
ID	& Description &  MAP \\ \hline
1	& level-1 models + \( \Gamma(1) \) ensemble& 0.4180 \\ \hline
2	& level-1 models + \( \Gamma(2) \) ensemble & 0.4252 \\ \hline
3	& level-1 models + \( \Gamma(3) \) ensemble & 0.4329 \\ \hline 
4	& level-1 models + \( \Gamma(4) \) ensemble & 0.4365 \\ \hline 
5	& level-1 models + \( \Gamma(5) \) ensemble & \textbf{0.4395} \\ \hline 
\end{tabular}
\label{tab2}
\end{table}

\section{Conclusions}
In this paper, we described the context-aware ensemble of multifaceted factorization models for recommendation prediction in social networks. SNS relationships and actions were integrated as implicit feedbacks. Various features, including age, gender, time, tweetnum,  keywords, tags, etc., were also used for user interest modeling.  A context-aware ensemble and user behavior modeling methods were proposed.  Experimental results on KDD-Cup data set demonstrated the effectiveness of the techniques proposed.


\section{Acknowledgments}
We would like to acknowledge the SIGKDD organizers for holding this exciting and inspiring competition. We also thank the colleagues at Shanda Innovation Institute for fruitful discussions. 

%
\bibliographystyle{abbrv}
\bibliography{kddcup-paper}  
%
%
\end{document}